\documentclass[a4paper]{jpconf}
\usepackage{graphicx}
\usepackage{amsmath,amssymb}
\def\be{\begin{equation}}
\def\ee{\end{equation}}
\def\ba{\begin{eqnarray}}
\def\ea{\end{eqnarray}}

\def\dd{{\rm d}}
\def\R{{\cal R}}
\renewcommand{\vec}[1]{\boldsymbol{#1}}
\def\lsim{\raise0.3ex\hbox{$\;<$\kern-0.75em\raise-1.1ex\hbox{$\sim\;$}}}
\def\gsim{\raise0.3ex\hbox{$\;>$\kern-0.75em\raise-1.1ex\hbox{$\sim\;$}}}

\def\theta{\vartheta}

\newcommand{\apj}{{Astrophys.\ J. }}

\begin{document}
\title{Anisotropic diffusion and the cosmic ray anisotropy}

\author{M.~Kachelrie\ss}

\address{Institutt for fysikk, NTNU, Trondheim, Norway}

\begin{abstract}
We  argue that the diffusion of cosmic rays in the Galactic
magnetic field  has to be strongly
anisotropic. As a result, the number of CR sources contributing
to the local CR flux is reduced by a factor $\sim 200$.
The CR density is therefore less
smooth, and the contribution of individual sources to the CR dipole
anisotropy becomes more prominent. In the case of anisotropic diffusion,
the observed plateau in the CR dipole anisotropy around 2--20\,TeV can be
explained by  a 2--3\,Myr old CR source which dominates the local CR flux
in this energy range. 

\end{abstract}

%%%%%%%%%%%%%%%%%%%%%%%%%%%%%%%%%%%%%%%%%%%%%%%%%%%%%%%%%%%%%%%%%%%%%%%%%%%
\section{Introduction}

The observed distribution of  cosmic ray (CR) arrival directions is
highly isotropic. Since Galactic CR sources are strongly concentrated in
the Galactic disc, an efficient mechanism for the isotropisation of the CR
momenta exists. Agent of this  isotropisation are turbulent magnetic fields,
since charged CRs scatter efficiently with resonant field modes which
wavelength matches their Larmor radius. As a result, CRs perform
on scales larger than the
coherence length of the turbulent field a random walk, and the memory
of the initial source location is mostly erased. Residual anisotropies
are connected to the structure of the local magnetic field and, e.g.,
to a remaining net flux of CRs.

Since large wavelengths of the turbulent field modes are less abundant,
CRs with higher energy are scattered less efficiently. Therefore, the
diffusion picture predicts that the CR anisotropy should increase
monotonically with energy. More precisely, if the turbulent field follows a
Kolmogorov power law as suggested by the observed B/C ratio, the dipole
anisotropy $\delta$ should increase with energy as $\delta\propto E^{1/3}$.
Both the energy-dependence
and the absolute value of the dipole anisotropy predicted in simple isotropic
diffusion models do not agree with observations. This discrepancy was dubbed
the ``CR anisotropy problem'' by Hillas \cite{Hillas:2005cs}.

In this short review based on the results of Refs.~\cite{Savchenko:2015dha,Giacinti:2017dgt}, we will first argue that the diffusion of CRs in the Galactic
magnetic field (GMF)
has to be strongly anisotropic. As a result, the number of CR sources
contributing
to the local CR flux is strongly reduced. Therefore, the CR density is less
smooth, and the contribution of individual sources to the CR dipole
anisotropy becomes more prominent than in the standard picture.
Then we argue that the observed
plateau in the CR dipole anisotropy around 2--20\,TeV is connected to a
2--3\,Myr old CR source which dominates in this energy range the local CR
flux. Finally, we comment on the alternative that a young source like Vela
is responsible for the observed plateau in the dipole anisotropy.

%%%%%%%%%%%%%%%%%%%%%%%%%%%%%%%%%%%%%%%%%%%%%%%%%%%%%%%%%%%%%%%%%%%%%%%%%%%

\section{Galactic magnetic field and anisotropic diffusion}

In the diffusion approach to CR propagation one considers  typically
the CR density in the stationary limit. The measured ratios of CR isotopes
like Be$^{10}$/Be$^{9}$ and of secondary/primary ratios  like B/C indicate a
residence time of CRs with rigidity $\R$ of order $\tau_{\rm esc}\simeq
{\rm few}\times 10^7{\rm yr}\left(\R/5\, {\rm GV}\right)^{-\beta}$ with
$\beta\simeq 1/3$. Then the flux from some $10^{4}$ sources accumulates at low
rigidities, forming a ``sea'' of Galactic CRs, if one assumes that the main
CR sources are supernovae (SN) injecting
$\simeq 10^{50}$\,erg every $\simeq 30$\,yr in the form of CRs.
Since many sources contribute, the discrete nature of the CR sources
can be neglected. Assuming additionally that the turbulent magnetic field
dominates
relative to the regular field, one often replaces the  diffusion tensor
$D_{ij}$ by a scalar diffusion coefficient $D$.

%%%%%%%%%%%%%%%%%%%%%%%%%%%%%%%%%%%%%%%%%%%%%%%%%%%
\begin{figure}
\begin{center}
  \includegraphics[width=0.45\columnwidth,angle=270]{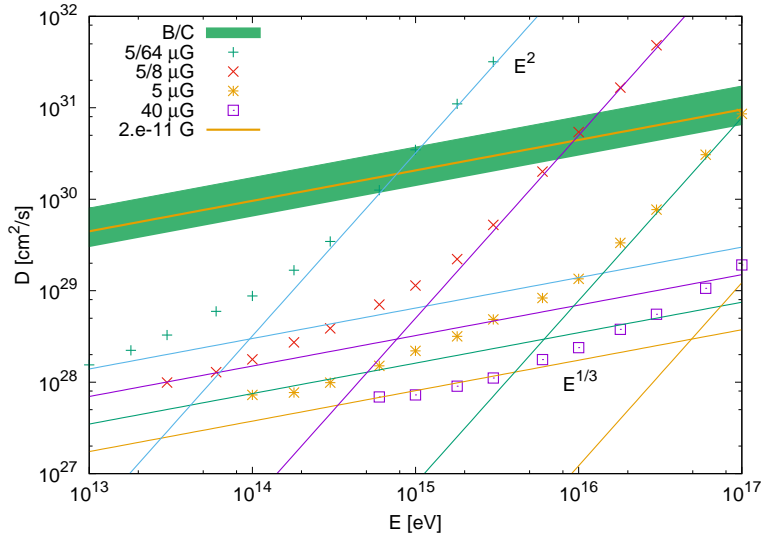}
\end{center}
\vskip0.6cm
\caption{CR diffusion coefficient $D(E)$ in pure isotropic Kolmogorov
  turbulence with $L_{\max}=25$\,pc for four values of the strength $B_{\rm rms}$
  of the turbulent field. 
 The green band shows the range of magnetic field strengths for which the diffusion coefficient satisfies $D_0=(3-8)\times 10^{28}$cm$^2$/s at $E_0=10$\,GeV;
 from Ref.~\protect\cite{Giacinti:2017dgt}.
\label{iso1}}
\end{figure}
%%%%%%%%%%%%%%%%%%%%%%%%%%%%%%%%%%%%%%%%%%%%%%%%%%%

The diffusion approach based on the approximations described above has
been sufficient to describe the bulk of experimental data obtained until
$\simeq 2005$. With the increased precision of newer experiments,
several discrepancies like the ``positron excess'' or breaks in the
CR spectra of nuclei have been emerged.
Here we will discuss a more theoretical challenge for the
approximations employed in the  standard diffusion approach,
which has also the potential to solve other observational
anomalies.

\begin{figure}
\begin{center}
\includegraphics[width=0.6\columnwidth]{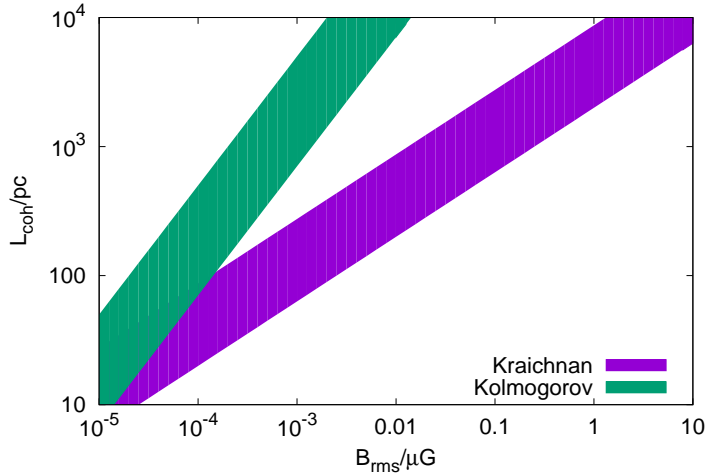}
\end{center}
\caption{Allowed ranges of $B_{\rm rms}$ and $L_{\rm coh}$ compatible
with $D_0=(3-8)\times 10^{28}$cm$^2$/s at $E_0=10$\,GeV for
Kolmogorov and Kraichnan turbulence. These ranges should be compared with 
the typical order-of-magnitude values that are relevant for the Galactic 
magnetic field: $B_{\rm rms} \sim (1 - 10)\,\mu$G and 
$L_{\rm coh} \lsim$~a few tens of pc, from Ref.~\protect\cite{Giacinti:2017dgt}.
\label{iso2}}
\end{figure}

In Fig.~\ref{iso1}, we  show the diffusion coefficient 
\begin{equation}
D_{ij}= \lim_{t\to\infty}
\frac{1}{2Nt}\sum_{a=1}^N (x_i^{(a)}-x_{i,0})(x_j^{(a)}-x_{j,0})
\label{D_iso}
\end{equation}
calculated numerically following the trajectories $x_i^{(a)}(t)$ of $N$ CRs 
injected into a pure random field with a Kolmogorov power spectrum
with $L_{\max}=25$\,pc for various field strengths.
The transition at $R_L(E_{\rm cr})=L_{\max}$ between the 
asymptotic low-energy ($D\propto E^{1/3}$, large-angle scattering) and 
high-energy ($D\propto E^{2}$, small-angle scattering) 
behaviour is clearly visible. However, for all used field strengths  
the diffusion coefficients are much smaller than those extracted using
e.g.\ Galprop~\cite{Johannesson:2016rlh} or DRAGON~\cite{Evoli:2008dv}.
Therefore CR propagation cannot be isotropic, because otherwise
CRs overproduce secondary nuclei like boron for any reasonable values of the
strength and the coherence scale of the turbulent field, cf.\ with
Fig.~\ref{iso2}. Such an anisotropy may appear if
the turbulent field at the considered scale does not dominate over the 
ordered component, or if the turbulent field itself is anisotropic.

\begin{figure}
  \includegraphics[width=0.45\columnwidth]{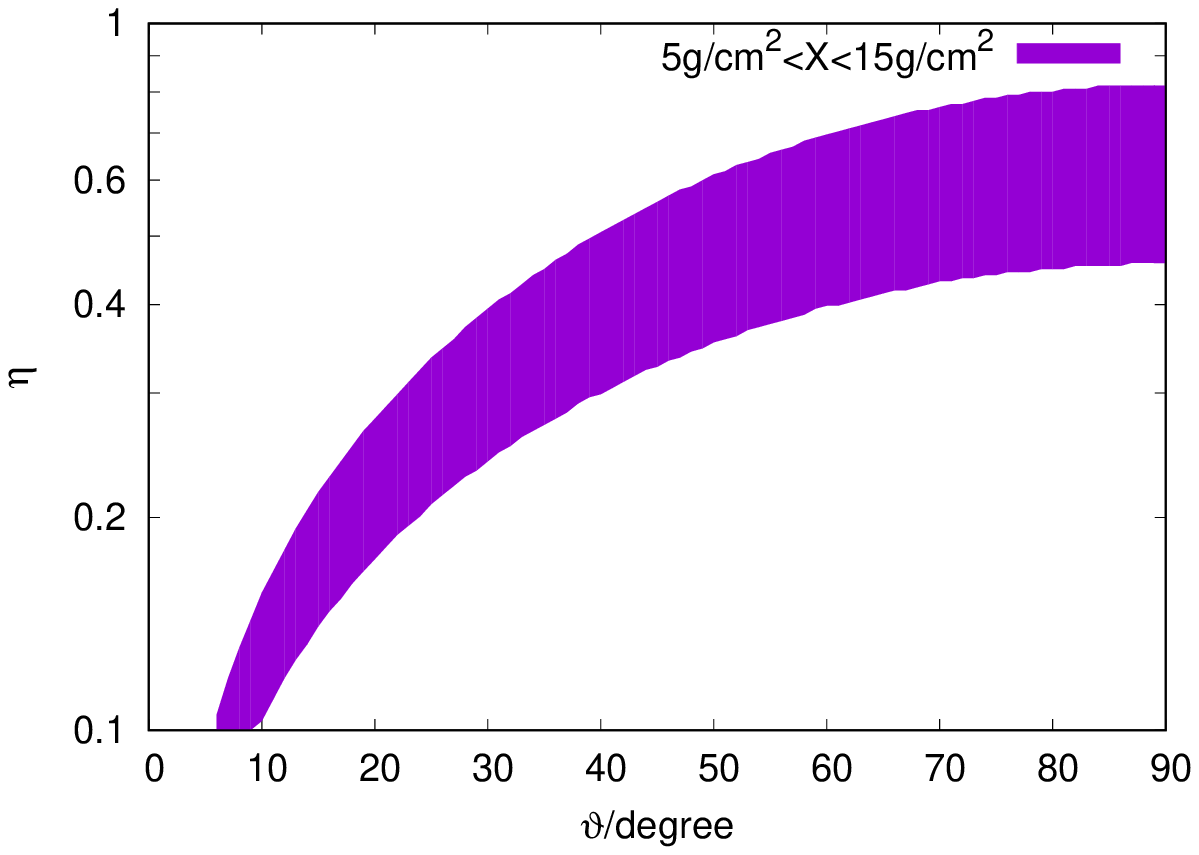}
  \includegraphics[width=0.45\columnwidth]{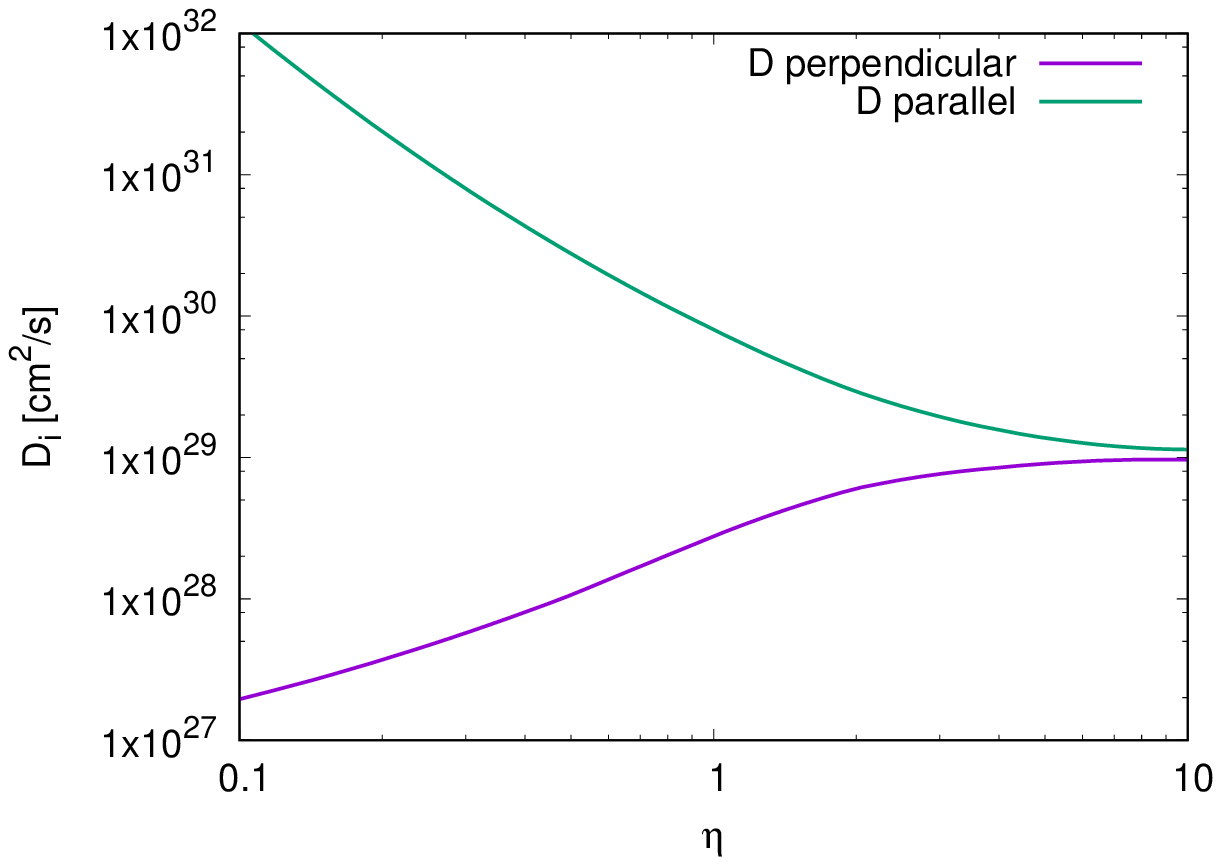}
  \caption{
    Left: Grammage $X$ crossed by CRs in a \lq\lq disc and halo\rq\rq\/ model as a function of the tilt angle $\theta$ between the regular magnetic field and the Galactic plane, and of the turbulence level $\eta$.
    Right: Fit of the diffusion coefficients $D_\|$ and $D_\perp$
    at $E=10^{15}$\,eV as a function of $\eta$,
    from Ref.~\protect\cite{Giacinti:2017dgt}.
\label{fig:Xeta}}
\end{figure}

One can estimate the level of anisotropy required considering the following
toy model: Let us adopt a thin matter disc with density
$\rho/m_{\rm p}\simeq 1$/cm$^3$ and height $h=150$\,pc around the Galactic
plane, while CRs propagate inside a larger halo of height $H=5$\,kpc. We assume
that the
regular magnetic field inside this disc and halo has a tilt angle $\theta$
with the Galactic plane, so that the component of the diffusion tensor
relevant for CR escape is given by
\be
D_z = D_\perp\cos^2\theta + D_\|\sin^2\theta \,.
\ee
Applying a simple leaky-box approach, the grammage follows as
$X = c \rho hH /D_z$.
Using now as allowed region for the grammage  $5\leq X\leq 15$\,g/cm$^2$,
the permitted region in the $\theta$--$\eta$ plane shown in the left panel of
Fig.~\ref{fig:Xeta} follows, where $\eta\equiv B_{\rm rms}/B_0$ describes the
turbulence level. For not too large values of the tilt angle,
$\theta\lsim 30^\circ$, the regular field should strongly dominate,
$\eta\lsim 0.35$. This results in a strongly  anisotropic propagation of CRs,
where the diffusion coefficient perpendicular to the ordered field can be
between two and three  orders of magnitude smaller than the parallel one,
$D_\bot\ll D_{||}$. As a result, the  $z$ component of the regular magnetic
field can drive CRs efficiently out of the Galactic disk. For instance, the
``X-field'' in the Jansson-Farrar model~\cite{Jansson:2012rt} for the GMF
leads to the correct CR escape time, if one chooses
$\eta\simeq 0.25$~\cite{Giacinti:2014xya}.

For this choice, the diffusion coefficients satisfy $D_\|\simeq 5 D_{\rm iso}$
and $D_\perp\simeq D_{\rm iso}/500$, where $D_{\rm iso}$ denotes 
the isotropic diffusion coefficient
$D_{\rm iso}$ satisfying the B/C constraints. In the regime, where the
CRs emitted by a single source fill a Gaussian with volume
$V(t)=\pi^{3/2} D_\perp D_\|^{1/2}t^{3/2}$, the CR density is increased
by a factor $500/\sqrt{5}\simeq 200$ compared to the case of isotropic
diffusion. The smaller volume occupied by CRs from each single source
leads to a smaller number 
of sources contributing substantially to the local flux, with only 
$\sim 10^2$ sources at $\R\sim 10$\,GV and about $\sim 10$~most recent SNe in 
the TeV range. This reduction of the effective number of sources 
may invalidate the assumption of a continuous CR injection and a stationary
CR flux.

%%%%%%%%%%%%%%%%%%%%%%%%%%%%%%%%%%%%%%%%%%%%%%%%%%%
%%%%%%%%%%%%%%%%%%%%%%%%%%%%%%%%%%%%%%%%%%%%%%%%%%%

%%%%%%%%%%%%%%%%%%%%%%%%%%%%%%%%%%%%%%%%%%%%%%%%%%%
%%%%%%%%%%%%%%%%%%%%%%%%%%%%%%%%%%%%%%%%%%%%%%%%%%%

%%%%%%%%%%%%%%%%%%%%%%%%%%%%%%%%%%%%%%%%%%%%%%%%%%%%%%%%%%%%%%%%%%%%%%%%%%% 
\section{A local source and the cosmic ray anisotropy}

In the diffusion approximation,  Fick's law is valid and the net CR current
$\vec j(E)$ is determined by the gradient of the CR number density 
$n(E)=\dd N/(\dd E \dd V)$ and the diffusion tensor $D_{ab}(E)$ as
$j_a = -D_{ab}\nabla_b n$. The dipole vector $\vec \delta$ of the CR
intensity $I= c/(4\pi)n$ follows then as
\begin{equation} \label{delta_diff}
 \delta_a 
          =  \frac{3}{c} \frac{j_a}{n} 
          = - \frac{3D_{ab}}{c}\frac{\nabla_b n}{n} \,.
\end{equation}
In the case of a strong ordered magnetic field $\vec B$, the tensor structure
of the diffusion tensor simplifies to $D_{ab}\propto B_aB_b$. This corresponds
to a projection of the CR gradient onto the magnetic field
direction~\cite{1990ApJ...361..162J}.
Hence, anisotropic diffusion predicts that the dipole anisotropy should
align with the local ordered magnetic field instead of pointing to the
source~\cite{1990ApJ...361..162J,Savchenko:2015dha}. Note that the ordered
magnetic field corresponds to the sum of the regular
magnetic field and the sum of turbulent field modes with wavelengths larger
than the Larmor radius at the corresponding CR energy.

In the case of an (anisotropic) three-dimensional Gaussian CR density $n$, the
formula~(\ref{delta_diff}) can be evaluated analytically. The result
$\delta=3R/(2cT)$ for a single source with age $T$ and distance
$R$ is independent of the regular and turbulent magnetic field.
In Ref.~\cite{Savchenko:2015dha}, it was shown that the CR density of
a single source is quasi-Gaussian, if CRs propagate over length scales
$l\gg L_{\rm coh}$. Numerically,  the dipole anisotropy $\delta$ of a
source  contributing the fraction $f_i$ to the total observed CR
flux is thus
\be \label{single}
 \delta_i = f_i \,\frac{3R}{2cT} \simeq
  5.0 \times 10^{-4} \:f_i\,\left(\frac{R}{200\,{\rm pc}}\right)
 \left(\frac{T}{2\,{\rm Myr}}\right)^{-1} \,.
\ee
%

%%%%%%%%%%%%%%%%%%%%%%%%%%%%%%%%%%%%%%%%%%%%%%%%%%%
\begin{figure}
\begin{center}
  \includegraphics[width=0.5\columnwidth,angle=270]{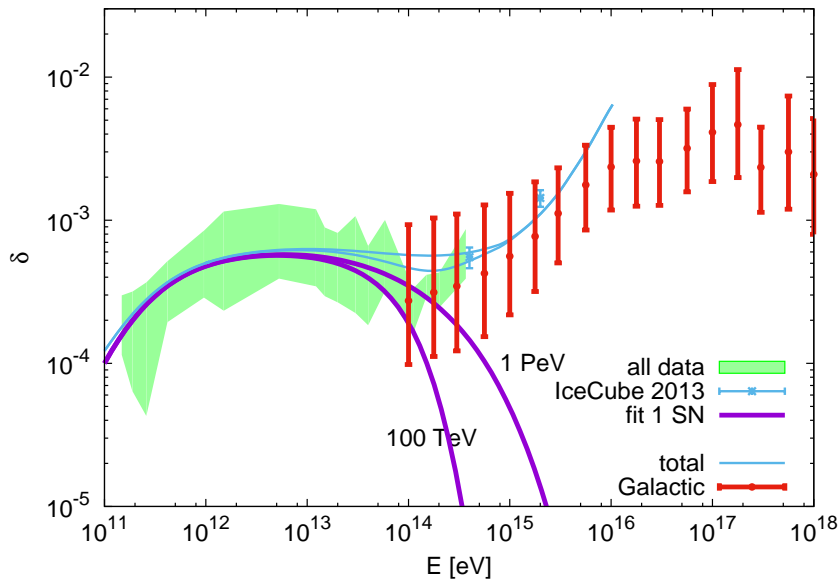}
\end{center}
\caption{Lower and upper limit (green band) 
on the dipole  anisotropy and data from IceCube (blue error-bars) compared 
to the contribution from the local  source (red, for two values of $E_{\max}$) 
and from the average CR sea  (magenta)  as function of energy; adapted
from Ref.~\protect\cite{Savchenko:2015dha}.
\label{Adata}}
\end{figure}
%%%%%%%%%%%%%%%%%%%%%%%%%%%%%%%%%%%%%%%%%%%%%%%%%%%

In Fig.~\ref{Adata}, we show experimental data for the
dipole anisotropy from Refs.~\cite{DiSciascio:2014jwa,Ahlers:2016rox}
as a green band:
The anisotropy grows as function of energy until $E\simeq 2$\,TeV,
remains approximately constant in the range 2--20\,TeV,
before it decreases again. The plateau in the range 2--20\,TeV is
naturally explained by the energy-independent contribution to the dipole
anisotropy of a single source. This is supported by the fact that the dipole
phase remains approximately constant in this range too, before it flips
by $\sim 180^\circ$.  Such a flip is naturally explained by  the projection
effect on the magnetic field line, if above 20\,TeV another source, which
is located in the opposite hemisphere, dominates the CR dipole anisotropy.

More specifically, it was suggested in Ref.~\cite{Savchenko:2015dha}
that a 2--3\,Myr old source at the distance 200--300\,pc dominates
the dipole anisotropy in the range 2--20\,TeV. Previously, it was
shown in Ref.~\cite{Kachelriess:2015oua} that the same source can explain
the ``positron excess'', as well as the breaks of the nuclei spectra
and the different slope of the proton spectrum~\cite{Kachelriess:2017yzq}.
The contribution of this local source is shown by two magenta lines for
two different high-energy cutoffs: In one case, it was assumed that the
source can accelerate up 100\,TeV, in the other that it is a PeVatron.
In both cases, the CR flux was calculated following the trajectories of
individual CRs, as discussed in Refs.~\cite{Giacinti:2014xya,Giacinti:2015hva,Kachelriess:2015oua,Kachelriess:2017yzq}.
Additionally, the total anisotropy beyond $10^{14}$\,eV of all Galactic SNe
is shown by red error-bars which is calculated in the escape model which uses
the same magnetic field configuration as the one used for the loal
source~\cite{Giacinti:2014xya,Giacinti:2015hva}.

A characteristic feature of this proposal is that a relatively old source
dominates the observed CR flux. This is only possible in the case of
anisotropic diffusion, and requires additionally that the perpendicular
distance $d_\perp$ of the Sun to the magnetic field line connecting it to
the source is not too large. Even for small $d_\perp$, the CR flux from the
single source is suppressed at low-energies, because of the slower
perpendicular diffusion.
In Refs.~\cite{Kachelriess:2017yzq}, the value $d_\perp\simeq 70$\,pc was
estimated requiring that the low-energy break in the source spectrum
explains the breaks in the energy spectra of CR nuclei. For this choice of
$d_\perp$,
the flux of the local source is suppressed below $\simeq 1$\,TeV (cf.\ with
Fig.~2 of Ref.~\cite{Kachelriess:2017yzq}), leading to a decreasing $f_i$
and the transition to the standard $\delta\propto E^{1/3}$ behavior below
this energy.

Another choice for the age of the source was suggested in Refs.~\cite{2013APh....50...33S,Ahlers:2016njd}.
Here, Vela with the age around 11,000\,yr and distance 300\,pc was
proposed as the single source responsible for the plateau in the
dipole anisotropy in the energy range 2--20\,TeV. In this case, the
contribution of Vela to the dipole amplitude has to be suppressed by a factor
$\simeq 200$.
Three mechanisms for such a suppression may be operating:  First, if
the regular magnetic field and the CR gradient are not parallel,
the projection effect in $D_{ab}\nabla_b n$ can reduce the
dipole~\cite{Mertsch:2014cua,Ahlers:2016njd}.
Second, the measured CR dipole is a projection into the equatorial
plane and is thus reduced compared to the true one.
Finally, the CR flux contributed by Vela may be small.
Calculating the CR fluxes from nearby young sources using the standard
isotropic diffusion coefficient and taking into account these effects,
Ref.~\cite{Ahlers:2016njd} argued that Vela leads to correct level of
anisotropy.
There is however a caveat in this conclusion: 
While Ref.~\cite{Ahlers:2016njd} calculates the CR fluxes from individual
sources using an {\em isotropic\/} diffusion coefficient, the remaining
analysis is based on the assumption of strongly anisotropic diffusion.
In the latter case, the CR flux depends
however crucially on the  perpendicular distance $d_\perp$ of the source to the
magnetic field line connecting it with the Sun, and a calculation of the
CR flux following the lines of Refs.~\cite{Savchenko:2015dha,Kachelriess:2015oua,Kachelriess:2017yzq} is required. Moreover, the number of sources is strongly
reduced and correspondingly the flux of nearby sources with small perpendicular
distance strongly enhanced.

%%%%%%%%%%%%%%%%%%%%%%%%%%%%%%%%%%%%%%%%%%%%%%%%%%%%%%%%%%%%%%%%%%%%%%%%%%% 
\section{Conclusions}

We have argued that the diffusion in the GMF has to be strongly anisotropic,
because otherwise CRs overproduce secondary nuclei like boron for any
reasonable values of the strength and the coherence scale of the turbulent
field. Therefore the number of CRs contributing to the local CR flux is
strongly reduced compared to the ``standard picture''. As a result, the CR
density is less smooth, and the contribution of individual sources to the
CR dipole anisotropy becomes more prominent. In this picture, the observed
plateau in the CR dipole anisotropy around 1--20\,TeV can be explained by a
2--3\,Myr old CR source which dominates the local CR flux in this energy range.
Such a source can explain also several other CR puzzles such as the
``positron excess'' , the difference in the slope of the proton and nuclei
spectra as well as their breaks~\cite{Kachelriess:2017yzq}.

%%%%%%%%%%%%%%%%%%%%%%%%%%%%%%%%%%%%%%%%%%%%%%%%%%%%%%%%%%%%%%%%%%%%%%%%%%%
\subsection*{Acknowledgments}
\noindent
It is a pleasure to thank Gwenal Giacinti, Andrii Neronov, Volodymyr Savchenko
and Dimitri Semikoz for fruitful collaborations.

%\bibliography{cr}

\begin{thebibliography}{10}
\expandafter\ifx\csname url\endcsname\relax
  \def\url#1{{\tt #1}}\fi
\expandafter\ifx\csname urlprefix\endcsname\relax\def\urlprefix{URL }\fi
\providecommand{\eprint}[2][]{\url{#2}}

\bibitem{Hillas:2005cs}
Hillas A~M 2005 {\em J. Phys.\/} {\bf G31} R95--R131

\bibitem{Savchenko:2015dha}
Savchenko V, Kachelrie{\ss} M and Semikoz D~V 2015 {\em Astrophys. J.\/} {\bf
  809} L23 (\textit{Preprint} \eprint{1505.02720})

\bibitem{Giacinti:2017dgt}
Giacinti G, Kachelrie{\ss} M and Semikoz D~V 2018 {\em JCAP\/} {\bf 1807} 051
  (\textit{Preprint} \eprint{1710.08205})

\bibitem{Johannesson:2016rlh}
J{\'o}hannesson G {\em et~al.\/} 2016 {\em Astrophys. J.\/} {\bf 824} 16
  (\textit{Preprint} \eprint{1602.02243})

\bibitem{Evoli:2008dv}
Evoli C, Gaggero D, Grasso D and Maccione L 2008 {\em JCAP\/} {\bf 0810} 018
  [Erratum: JCAP1604,no.04,E01(2016)] (\textit{Preprint} \eprint{0807.4730})

\bibitem{Jansson:2012rt}
Jansson R and Farrar G~R 2012 {\em Astrophys.J.\/} {\bf 761} L11
  (\textit{Preprint} \eprint{1210.7820})

\bibitem{Giacinti:2014xya}
Giacinti G, Kachelrie{\ss} M and Semikoz D~V 2014 {\em Phys. Rev.\/} {\bf D90}
  041302 (\textit{Preprint} \eprint{1403.3380})

\bibitem{1990ApJ...361..162J}
{Jones} F~C 1990 {\em \apj\/} {\bf 361} 162--172

\bibitem{DiSciascio:2014jwa}
Di~Sciascio G and Iuppa R 2014  (\textit{Preprint} \eprint{1407.2144})

\bibitem{Ahlers:2016rox}
Ahlers M and Mertsch P 2017 {\em Prog. Part. Nucl. Phys.\/} {\bf 94} 184--216
  (\textit{Preprint} \eprint{1612.01873})

\bibitem{Kachelriess:2015oua}
Kachelrie{\ss} M, Neronov A and Semikoz D~V 2015 {\em Phys. Rev. Lett.\/} {\bf
  115} 181103 (\textit{Preprint} \eprint{1504.06472})

\bibitem{Kachelriess:2017yzq}
Kachelrie{\ss} M, Neronov A and Semikoz D~V 2018 {\em Phys. Rev.\/} {\bf D97}
  063011 (\textit{Preprint} \eprint{1710.02321})

\bibitem{Giacinti:2015hva}
Giacinti G, Kachelrie{\ss} M and Semikoz D~V 2015 {\em Phys. Rev.\/} {\bf D91}
  083009 (\textit{Preprint} \eprint{1502.01608})

\bibitem{2013APh....50...33S}
{Sveshnikova} L~G, {Strelnikova} O~N and {Ptuskin} V~S 2013 {\em Astroparticle
  Physics\/} {\bf 50} 33--46 (\textit{Preprint} \eprint{1301.2028})

\bibitem{Ahlers:2016njd}
Ahlers M 2016 {\em Phys. Rev. Lett.\/} {\bf 117} 151103 (\textit{Preprint}
  \eprint{1605.06446})

\bibitem{Mertsch:2014cua}
Mertsch P and Funk S 2015 {\em Phys. Rev. Lett.\/} {\bf 114} 021101
  (\textit{Preprint} \eprint{1408.3630})

\end{thebibliography}

\section*{References}

\providecommand{\newblock}{}

\end{document}